\pdfoutput=1

\documentclass[11pt,twoside,a4paper,cmspaper,final,collab]{cms-tdr}
\def\svnVersion{177388}\def\cmsCernNo{2012-228}\def\cmsCernDate{\today}\def\cmsMessage{Submitted to Physical Review Letters}

\begin{document}\cmsNoteHeader{HIN-11-011}

\hyphenation{had-ron-i-za-tion}
\hyphenation{cal-or-i-me-ter}
\hyphenation{de-vices}
\RCS$HeadURL: svn+ssh://svn.cern.ch/reps/tdr2/papers/HIN-11-011/trunk/HIN-11-011.tex $
\RCS$Id: HIN-11-011.tex 150141 2012-10-01 19:36:20Z nuno $
\providecommand{\HYDJET} {{\textsc{hydjet}}\xspace}
\providecommand{\CASCADE} {{\textsc{cascade}}\xspace}
\providecommand{\linenomath}{\relax}
\newcommand{\pp}{{\ensuremath{\Pp\Pp}}\xspace}
\newcommand{\PbPb}{\ensuremath{\text{PbPb}}\xspace}
\newcommand{\sqrts}{\ensuremath{\sqrt{s}}\xspace}
\newcommand{\sqrtsnn}{\ensuremath{\sqrt{s_{NN}}}\xspace}
\newcommand{\PgUn}{\ensuremath{\Upsilon\text{(nS)}}\xspace}
\newcommand{\raa}{\ensuremath{R_\mathrm{AA}}\xspace}
\newcommand{\taa}{\ensuremath{T_\mathrm{AA}}\xspace}
\newcommand {\npart}  {\ensuremath{N_{\text{part}}}\xspace}
\newcommand {\ncoll}  {\ensuremath{N_{\text{coll}}}\xspace}
\renewcommand{\PgU}{\ensuremath{\Upsilon}\xspace}
\renewcommand{\PgUa}{\ensuremath{\Upsilon\text{(1S)}}\xspace}
\renewcommand{\PgUb}{\ensuremath{\Upsilon\text{(2S)}}\xspace}
\renewcommand{\PgUc}{\ensuremath{\Upsilon\text{(3S)}}\xspace}
\newcommand{\PgUbc}{\ensuremath{\Upsilon\text{(2S+3S)}}\xspace}
\ifthenelse{\boolean{cms@external}}{\providecommand{\cmsLeft}{top}}{\providecommand{\cmsLeft}{left}}
\ifthenelse{\boolean{cms@external}}{\providecommand{\cmsRight}{bottom}}{\providecommand{\cmsRight}{right}}
\ifthenelse{\boolean{cms@external}}{\providecommand{\suppMaterial}{the supplemental material [URL will be inserted by publisher]}}{\providecommand{\suppMaterial}{App.~\ref{app:suppMat}}}
\cmsNoteHeader{HIN-11-011}
\title{Observation of sequential Upsilon suppression in PbPb collisions}

\date{\today}

\abstract{The suppression of the individual \PgUn states in \PbPb collisions with respect to their yields in \pp data has been measured. The \PbPb and \pp datasets used in the analysis correspond to integrated luminosities of 150\mubinv and 230\nbinv, respectively, collected in 2011 by the CMS experiment at the LHC, at a center-of-mass energy per nucleon pair of 2.76\TeV. The \PgUn yields are measured from the dimuon invariant mass spectra.  The suppression of the \PgUn yields in \PbPb relative to the yields in \pp scaled by the number of nucleon-nucleon collisions, \raa, is measured as a function of the collision centrality. Integrated over centrality, the \raa values are $0.56 \pm 0.08\,\text{(stat.)} \pm 0.07\,\text{(syst.)}$, $0.12 \pm 0.04\,\text{(stat.)} \pm 0.02\,\text{(syst.)}$, and lower than 0.10 (at 95\% confidence level), for the \PgUa, \PgUb, and \PgUc states, respectively. The results demonstrate the sequential suppression of the \PgUn states in \PbPb collisions at LHC energies.}

\hypersetup{%
pdfauthor={CMS Collaboration},%
pdftitle={Observation of sequential Upsilon suppression in PbPb collisions},%
pdfsubject={CMS},%
pdfkeywords={physics, dimuons, heavy ion, bottomonia, suppression, quark gluon plasma}}

\maketitle

Suppression of heavy quarkonium states has been proposed as a probe of the properties of
the hot and dense medium created in high-energy heavy-ion collisions~\cite{Matsui:1986dk}.
If a deconfined state, often referred to as the quark-gluon plasma (QGP),
is formed, the confining potential of heavy quark-antiquark pairs is expected
to be screened because of interactions with quarks and gluons in the medium.
The resulting dissociation of the quarkonium states depends on
the temperature of the medium, and is expected to occur sequentially,
reflecting the increasing values of their binding energies~\cite{Digal:2001ue}.
The $\PgUa$ is the most tightly bound quarkonium state, and is hence expected
to be the one with the highest dissociation temperature.

The prediction of the suppression pattern is complicated by various factors.
These include feed-down contributions from higher-mass resonances into the observed quarkonium yields,
as well as several competing nuclear and medium effects.
These factors have played an important role in the interpretation of the charmonium measurements~\cite{Zhao:2011cv}.
The bottomonium family is expected to provide additional
and theoretically cleaner probes of the deconfined medium.
The three $\PgUn$ states, characterized by similar decay kinematics
but distinct binding energies, further enable the measurement of relative state
suppression, where common experimental and theoretical factors, and respective uncertainties, cancel.

Measurements of the absolute $\PgUa$ suppression~\cite{HIN-10-006} and of the relative suppression of $\PgUb + \PgUc$ with respect to $\PgUa$~\cite{prl} were recently reported. These analyses used \PbPb (\pp) data corresponding to an integrated luminosity of $7.3 \mubinv$ ($230 \nbinv$)
collected in 2010 (2011) at the same center-of-mass energy per nucleon pair of $\sqrtsnn =$ 2.76\TeV,
 with the Compact Muon Solenoid (CMS) detector at the Large Hadron Collider (LHC).
The total \PgU yields in \PbPb (\pp) collisions are denoted by $\PgU|_{\PbPb}$ ($\PgU|_{\pp}$).
Selecting reconstructed muons with pseudorapidity $|\eta|<2.4$ and transverse momentum $\pt > 4$\GeVc,
the $\PgUa$ nuclear modification factor
\raa (defined in Eq.~(\ref{eqn:raa}))
was measured to be
$0.63 \pm 0.11\,\text{(stat.)} \pm 0.10\,\text{(syst.)}$.
The double ratio
$\frac{\PgU\text{(2S+3S)}/\PgUa|_{\PbPb}}{\PgU\text{(2S+3S)}/\PgUa|_{\pp}}$
was measured in the same muon kinematic region to be
$0.31 _{-0.15}^{+0.19}\,\text{(stat.)} \pm 0.03\,\text{(syst.)}$, indicating
that the excited \PgUn\ states are suppressed with respect to the \PgUa,
at a significance of 2.4 standard deviations ($\sigma$).
In this Letter, an update of these measurements is reported,
utilizing a \PbPb data sample corresponding to an integrated luminosity of $150 \mubinv$, collected in 2011 by CMS,
 at $\sqrtsnn =$ 2.76\TeV as in the previous study.
This larger \PbPb dataset together with the excellent momentum resolution of the CMS detector enables the separation of all three $\PgU$ states below open-bottom threshold for the first time in the heavy-ion environment and the measurement of the centrality dependence of their yields.

A detailed description of the CMS detector can be found elsewhere~\cite{CMS:2008zzk}. Its central feature is a superconducting solenoid of 6\unit{m} internal diameter, providing a magnetic field of 3.8\unit{T}. Within the field volume are the silicon  pixel and strip tracker, the crystal electromagnetic calorimeter, and the brass/scintillator hadron calorimeter.
The silicon pixel and strip tracker measures charged-particle trajectories in the  range $|\eta| < 2.5$.
The tracker consists of 66\unit{M} pixel and 10\unit{M} strip sensor elements.
Muons are detected in the range $|\eta| < 2.4$, with detection planes based on three technologies: drift tubes, cathode strip chambers, and resistive plate chambers.
Because of the strong magnetic field and the fine granularity of the tracker, the muon \pt measurement based on information from the tracker alone has a resolution between 1 and 2\% for a typical muon in this analysis.

The CMS apparatus also has extensive forward calorimetry, including two steel/quartz-fiber \v{C}erenkov hadron forward calorimeters (HF), which cover the range $2.9 < |\eta| < 5.2$.
These detectors are used for event selection and centrality determination in \PbPb collisions.
The event centrality observable corresponds to the fraction of the total inelastic cross section, starting at 0\% for the most central collisions and
evaluated as percentiles of the distribution of the energy deposited in the HF~\cite{HIN-10-004,D'Enterria:2007xr}.
The centrality classes used in this analysis
are 50--100\%, 40--50\%, 30--40\%, 20--30\%, 10--20\%, 5--10\%, and 0--5\%,
ordered from the lowest to the highest HF energy deposit.
Using a Glauber-model calculation as described in Ref.~\cite{HIN-10-004},
the average number of nucleons participating in the collisions (\npart) and the average
nuclear overlap function (\taa) have been estimated for each centrality class.
The \taa factor is equal to the number of elementary
nucleon-nucleon (NN) binary collisions divided by the elementary NN
cross section and can be interpreted as the NN-equivalent integrated
luminosity per heavy-ion collision, at a given event centrality~\cite{Miller:2007ri}.

The $\PgU$ states are identified through their dimuon decay.
The events are selected online with a hardware-based trigger
requiring two muon candidates in the muon detectors.
More stringent muon quality requirements are imposed in the \PbPb case relative to the \pp online selection.
No explicit momentum or rapidity thresholds are applied at trigger level.
For the \PbPb data, events are preselected offline if they contain a reconstructed primary vertex comprising at least two tracks,
and the presence of energy deposits larger than 3\GeV in at least three towers in each of the two HF calorimeters.
These criteria reduce contributions from single-beam interactions,
 ultra-peripheral electromagnetic interactions, and cosmic-ray muons.

\begin{figure}[t]
\begin{center}
    \includegraphics[width=0.45\textwidth]{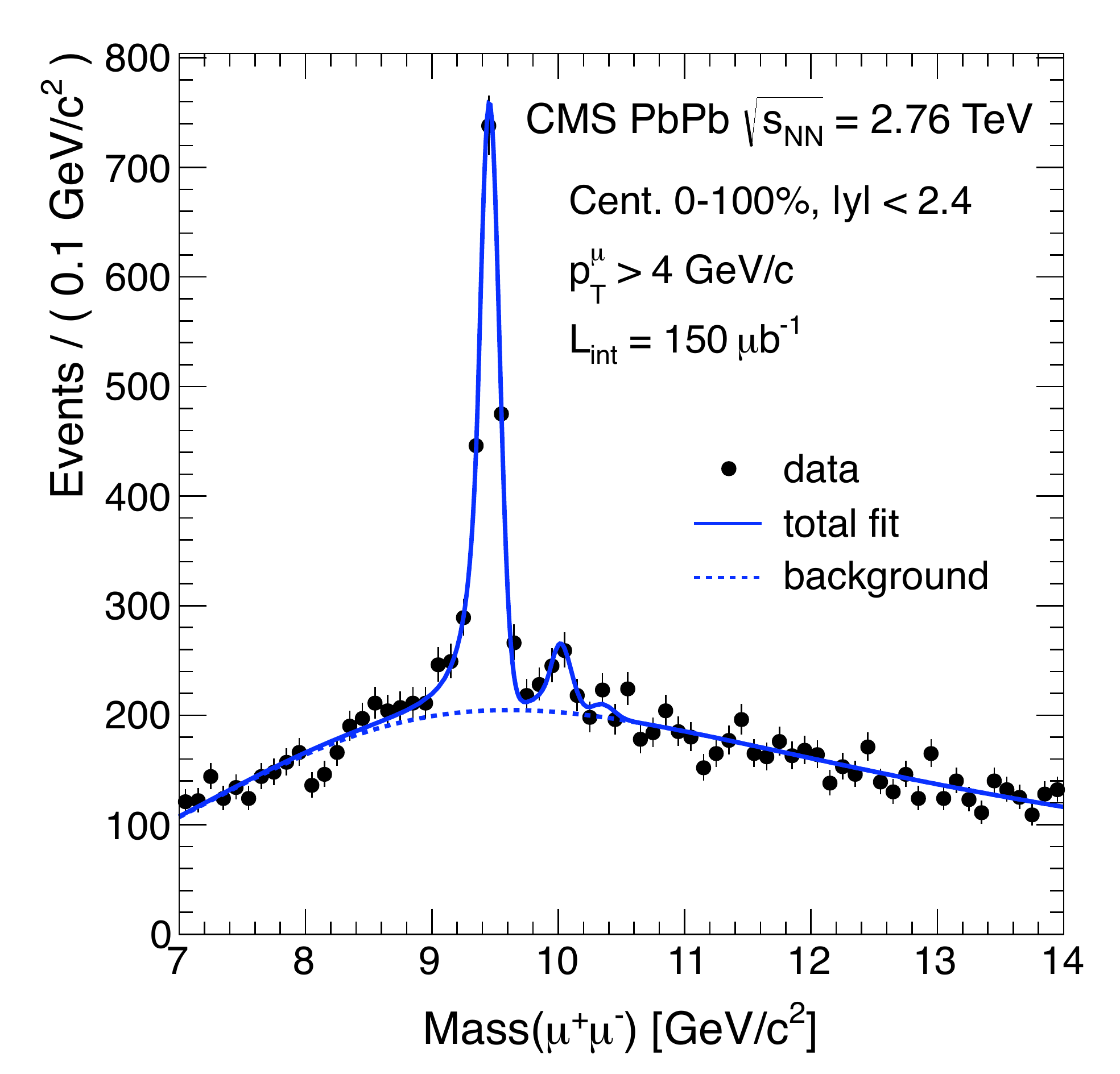}
    \includegraphics[width=0.45\textwidth]{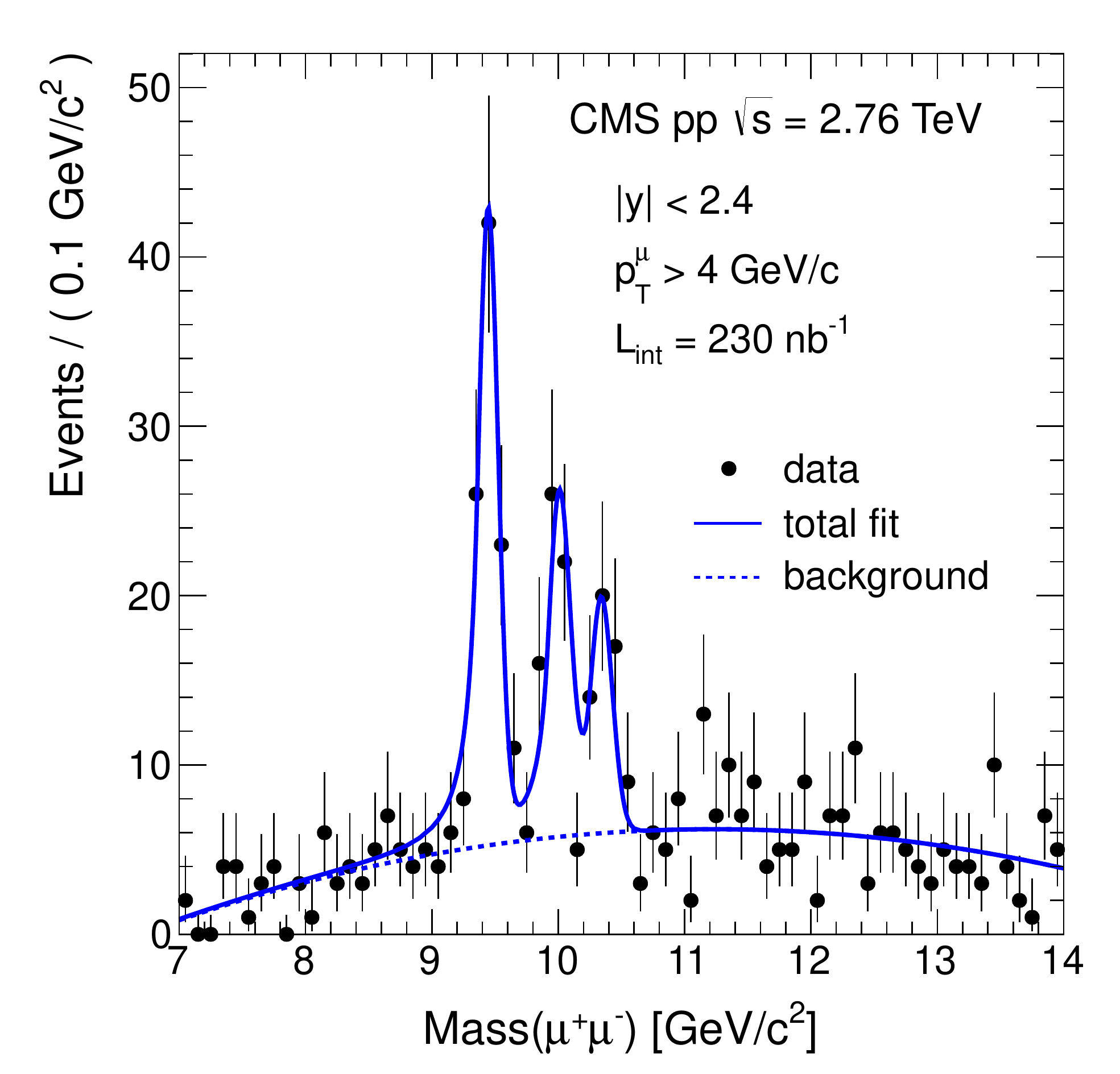}
    \caption{Dimuon invariant-mass distributions in \PbPb\ (\cmsLeft) and \pp (\cmsRight)
data at $\sqrtsnn = 2.76$\TeV. The same reconstruction algorithm and analysis selection are applied to both datasets, including a transverse momentum requirement on single muons of $\pt > 4\GeVc$. The solid (signal + background) and dashed (background-only) curves show the results of the simultaneous fit to the two datasets.}
    \label{fig:mass}
\end{center}
\end{figure}

Muons are reconstructed by matching tracks in the muon detectors and silicon tracker.
The same offline reconstruction algorithm and selection criteria are applied to the \PbPb and \pp data samples.
The muon candidates are required to have a transverse (longitudinal) distance of closest approach to the event vertex smaller than 3 (15)\cm.
Muons are only kept if the part of their trajectory in the tracker
has 11 or more hits and the $\chi^2$ per degree of freedom of the
combined and tracker-only fits are lower than 20 and 4, respectively.
Pairs of oppositely charged muons are considered dimuon candidates if the
$\chi^2$ fit probability of the tracks originating from a common vertex exceeds 5\%.
This removes background arising primarily from the displaced, semileptonic decays of charm and bottom hadrons.
Only muons with $\pt > 4$\GeVc and $|\eta|<2.4$ are considered, as in Ref.~\cite{prl}.
The dimuon \pt distribution of the selected candidates extends down to zero and has a mean of about 6\GeVc, 
covering a dimuon rapidity range of $|y|<2.4$.
The resultant dimuon invariant mass spectra are shown in Fig.~\ref{fig:mass} for the \PbPb and \pp datasets.  The three $\PgUn$ peaks are clearly observed in the \pp case; the $\PgUc$ state is not prominent above the dimuon continuum in \PbPb collisions.

Simulated Monte Carlo (MC) events are used to optimize muon selection cuts and to evaluate efficiencies.
Signal $\PgUn$ events are generated using \PYTHIA 6.424~\cite{Sjostrand:2006za},
with non-relativistic quantum chromodynamics matrix elements tuned
by comparison with CDF data~\cite{Acosta:2004yw}.
Underlying heavy-ion events are produced with the \HYDJET 1.6~\cite{Lokhtin:2005px} event generator.
The detector response is simulated with \GEANTfour~\cite{Agostinelli:2002hh}.
The signal candidates are embedded in the underlying \PbPb events,
at the level of detector hits and with matching vertices.
The resulting embedded events are then processed through the trigger emulation and the full event reconstruction chain.

An extended unbinned maximum likelihood fit to the two invariant mass spectra shown in Fig.~\ref{fig:mass}
is performed to extract the $\PgUn$ yields, following the method described in Refs.~\cite{prd, prl}. The measured mass lineshape of each $\PgUn$ state is parameterized by a ``Crystal Ball'' (CB) function, i.e.\ a Gaussian resolution function with the low-side tail replaced by a power law describing final-state radiation.
The mass differences between the states are fixed to their world average values~\cite{pdg2012} and the mass resolution is forced to scale with the resonance mass.
In our previous measurement~\cite{prl}, the signal shape parameters were fixed from MC simulation, including the mass resolution and CB tail parameters.
The current 20-fold larger \PbPb dataset allows these constraints to be released,
but the shape parameters are treated as common for both \PbPb and \pp datasets via a simultaneous fit.

The background model for the \pp dataset consists of a second-order polynomial, as was used in Ref.~\cite{prl},
while the larger \PbPb dataset requires a more detailed background model.
The $\pt > 4\GeVc$ muon selection threshold
causes a depletion of dimuon candidates in the lower part of the 7--14\GeVcc mass fitting range.
The \PbPb background model consists of an exponential function multiplied by an error function
describing the low-mass turn-on.
The background parameters are determined from the fit.
This nominal model accurately describes the mass sidebands in the opposite-sign muon signal sample, shown in
Fig.~\ref{fig:mass} (\cmsLeft),
as well as the alternative estimates of the shape of the combinatorial background obtained from like-sign muon pairs or via a ``track-rotation'' method. In the latter method~\cite{TrackRotation} the azimuthal angular coordinate of one of the muon tracks is rotated by 180 degrees.

The ratios of the observed yields, not corrected for differences in acceptance and efficiency, of the \PgUb\ and \PgUc\
states to the \PgUa\ state, in the \PbPb and \pp data, are
\begin{linenomath}
\begin{eqnarray}
\label{eqn:single-ratio}
  \PgUb/\PgUa|_{\pp}     & = & 0.56 \pm 0.13\,\text{(stat.)} \pm 0.02\,\text{(syst.)}\,,
\ifthenelse{\boolean{cms@external}}{\nonumber}{}\\
  \PgUb/\PgUa|_{\PbPb}   & = & 0.12 \pm 0.03\,\text{(stat.)} \pm 0.02\,\text{(syst.)}\,, \nonumber\\
  \PgUc/\PgUa|_{\pp}     & = & 0.41 \pm 0.11\,\text{(stat.)} \pm 0.04\,\text{(syst.)}\,, \nonumber \\
  \PgUc/\PgUa|_{\PbPb}   & = & 0.02 \pm 0.02\,\text{(stat.)} \pm 0.02\,\text{(syst.)} \nonumber
\ifthenelse{\boolean{cms@external}}{
\\ &&< 0.07 \text{ (95\% confidence level)} \,,
}{
\quad \left(< 0.07 \text{ at 95\% confidence level} \right) \,,
}
\end{eqnarray}
\end{linenomath}
where the systematic uncertainty arises from the fitting procedure, as described below.
For the \PgUc\ to \PgUa\ ratio in \PbPb, a 95\% confidence level (CL) limit is set,
based on the Feldman--Cousins statistical method~\cite{fc-stat}.

The measurement of the ratio of the $\PgUn/\PgUa$ ratios in \PbPb and \pp collisions benefits from an almost complete cancellation of possible acceptance or efficiency differences among the reconstructed resonances.
The simultaneous fit to the \PbPb and \pp mass spectra gives the double ratios
\begin{linenomath}
\begin{eqnarray}
\label{eqn:double-ratio}
  \frac{\PgUb/\PgUa|_{\PbPb}}{\PgUb/\PgUa|_{\pp}}
& = &  0.21 \pm 0.07\,(\text{stat.}) \pm 0.02\,(\text{syst.}) \,,
\ifthenelse{\boolean{cms@external}}{\nonumber}{}\\
  \frac{\PgUc/\PgUa|_{\PbPb}}{\PgUc/\PgUa|_{\pp}}
& = &  0.06 \pm 0.06\,(\text{stat.}) \pm 0.06\,(\text{syst.})  \nonumber
\ifthenelse{\boolean{cms@external}}{
\\ &&< 0.17 \text{ (95\% CL)} \,.
}{
\quad \left(< 0.17 \text{ at 95\% CL} \right) \,.
}
\label{eq:double}
\end{eqnarray}
\end{linenomath}
The systematic uncertainties from the fitting procedure are evaluated by varying the fit function as follows:
fixing the CB tail and resolution parameters to MC expectations,
allowing for differences in these parameters between \PbPb and \pp,
and constraining the background parameters with the like-sign and track-rotated spectra.
An additional systematic uncertainty (1\%), estimated from MC simulation, is included to account for possible imperfect cancellations of acceptance and efficiency.

The double ratios, defined in Eq.~(\ref{eqn:double-ratio}), are expected to be compatible with unity in the absence of
suppression of the excited states relative to the $\PgUa$ state.
The measured values are, instead, considerably smaller than unity.
The significance of the observed suppression exceeds $5\,\sigma$.

\begin{figure}[t]
\begin{center}
   \includegraphics[width=0.45\textwidth]{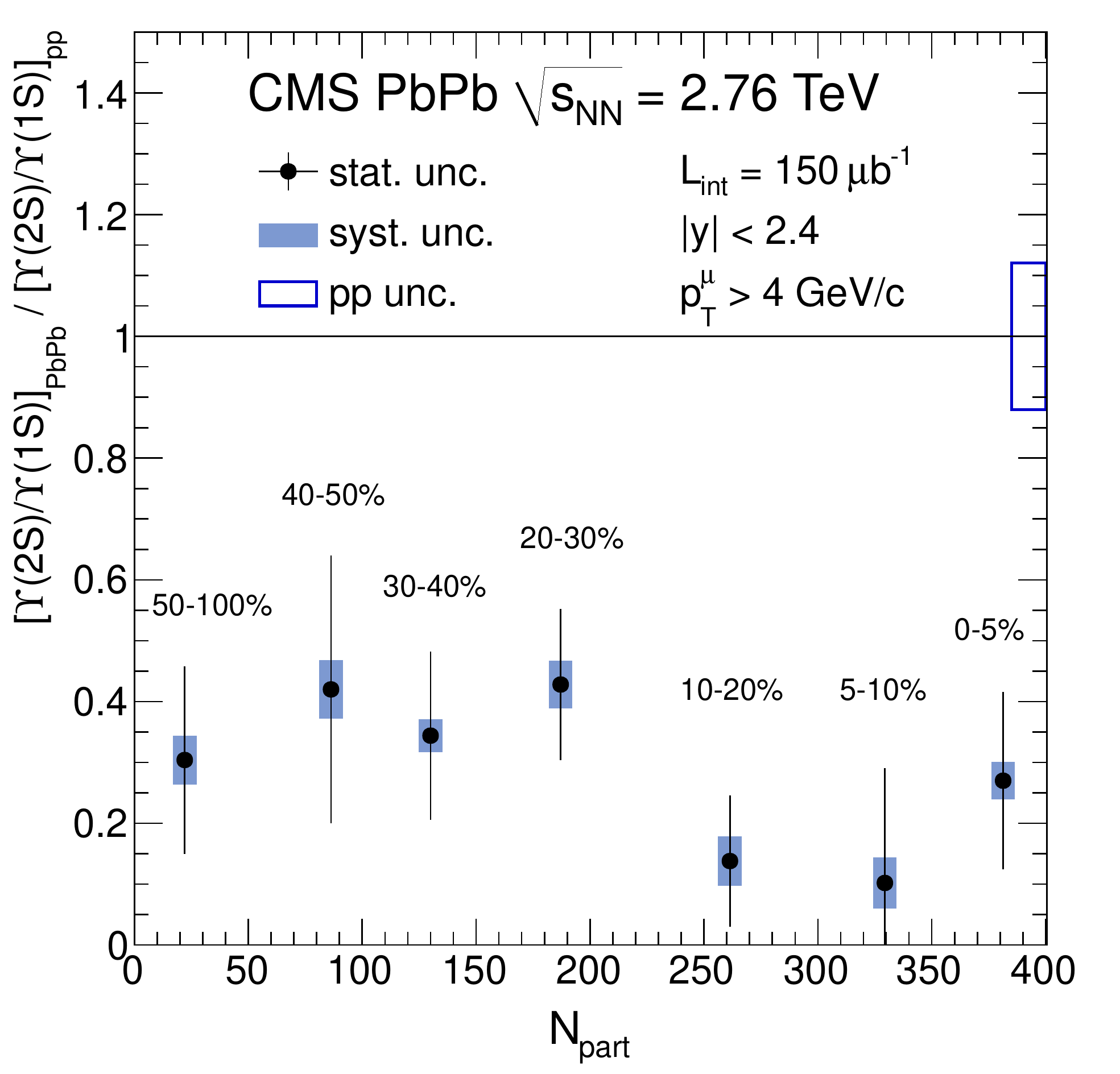}
   \includegraphics[width=0.45\textwidth]{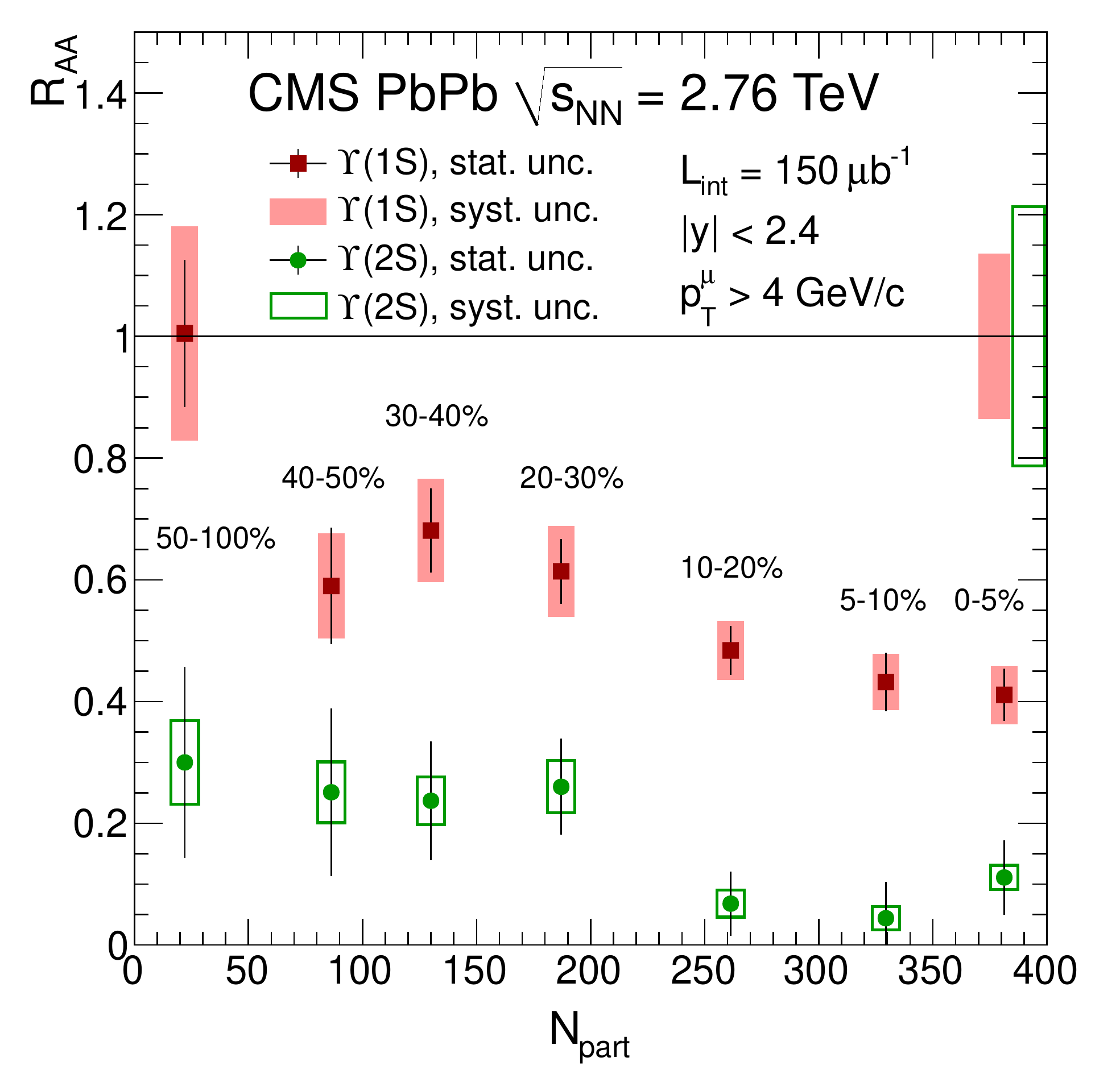}
  \caption{Centrality dependence of the double ratio (\cmsLeft) and of the nuclear modification factors (\cmsRight) for the $\PgUa$ and $\PgUb$ states. The relative uncertainties from \npart-independent quantities (\pp yields and, for the \raa, also integrated luminosity) are represented by the boxes at unity, and are not included in the data points as these uncertainties do not affect the point-to-point trend. The event centrality bins used are indicated by percentage intervals. The results are available in tabulated form in \suppMaterial.}
\label{fig:centrality}
\end{center}
\end{figure}

In order to investigate the dependence of the suppression on the centrality of the collision,
the double ratio $\frac{\PgUb/\PgUa|_{\PbPb}}{\PgUb/\PgUa|_{\pp}}$ is displayed as a function of \npart in
Fig.~\ref{fig:centrality}~(\cmsLeft).
The results are constructed from the single ratio $\PgUb/\PgUa|_{\PbPb}$
measured in bins of \PbPb centrality,  using the \pp ratio as normalization.
The dependence on centrality is not pronounced. 
More data, in particular more \pp collisions, are needed to establish possible
dependences on dimuon kinematic variables.

Absolute suppressions of the individual \PgU\ states and their dependence
on the collision centrality are studied using the nuclear modification factor, \raa,
defined as the yield per nucleon-nucleon collision in \PbPb relative to that in \pp.
The \raa observable,
\begin{linenomath}
\begin{equation}
\label{eqn:raa}
    \raa = \frac{\lumi_{\pp}}{\taa N_{\text{MB}}}\frac{\PgUn|_{\PbPb}}{\PgUn|_{\pp}} \frac{\varepsilon_{\pp}}{\varepsilon_{\PbPb}}\,,
\end{equation}
\end{linenomath}
is evaluated from the ratio of total $\PgUn$ yields in \PbPb and \pp collisions
corrected for the difference in efficiencies ${\varepsilon_{\pp}}/{\varepsilon_{\PbPb}}$,
with the average nuclear overlap function \taa,
number of minimum-bias events sampled by the event selection $N_{\mathrm{MB}}$,
and integrated luminosity of the \pp dataset  $\lumi_{\pp}$
accounting for the normalization.
The centrality-integrated (0--100\%) \raa\ values for the individual \PgU states are:
\begin{linenomath}
\begin{eqnarray}
\raa (\PgUa) &=& 0.56 \pm 0.08\,\text{(stat.)} \pm 0.07\,\text{(syst.)} \,,  \label{eqn:raa_results}
\ifthenelse{\boolean{cms@external}}{\nonumber}{}\\
\raa (\PgUb) &=& 0.12 \pm 0.04\,\text{(stat.)} \pm 0.02\,\text{(syst.)} \,, \nonumber \\
\raa (\PgUc) &=& 0.03 \pm 0.04\,\text{(stat.)} \pm 0.01\,\text{(syst.)}  \nonumber
\ifthenelse{\boolean{cms@external}}{
\\ &&< 0.10 \text{ (95\% CL)} \,.
}{
\quad \left(< 0.10 \text{ at 95\% CL} \right) \,.
}
\end{eqnarray}
\end{linenomath}
As the \PgUc\ peak is not prominent above the dimuon continuum (statistical significance less than one standard deviation), an upper limit is also presented.
The results for the \PgUa\ and \PgUb\ obtained by performing the measurement in ranges of centrality are displayed in
Fig.~\ref{fig:centrality}~(\cmsRight).

Each factor entering in Eq.~(\ref{eqn:raa}) contributes to the \raa uncertainty, including
$\lumi_{\pp}$ (6\%) and \taa (4--15\%, from central to peripheral collisions).
The systematic uncertainties from the fitting procedure, used in the determination of the $\PgUa$
(4--9\%), $\PgUb$ (10--40\%), and $\PgUc$ (14\%) signal yields, are estimated
as previously described for the double-ratio measurement.
The ratio of efficiencies in~Eq.~(\ref{eqn:raa}) is estimated from MC simulation to deviate by less than 7\% from unity for the centrality bins considered.
Systematic uncertainties on the efficiency ratio are estimated by considering variations of simulated kinematic distributions (5--7\%) and from differences in the efficiency ratio estimations from data and MC simulations (3\%).
For the former source, uncertainties are estimated by applying a weight to the generated \PgU\ \pt and $|y|$ distributions that increases linearly from 0.7 to 1.3 over the ranges  $0<\pt<20\GeVc$ and  $0<|y|<2.4$.
For the latter source, reconstruction and trigger selection efficiencies
are estimated employing a tag-and-probe method~\cite{prd,HIN-10-006},
using muons from \JPsi decays in \PbPb and \pp simulations as well as in collision data.

The results indicate a significant suppression of the $\PgUn$ states in heavy-ion collisions compared to \pp collisions at the same per-nucleon-pair energy.
The data support the hypothesis of increased suppression of less strongly bound states: the $\PgUa$ is the least suppressed and the $\PgUc$ is the most suppressed of the three states.
The $\PgUa$ and $\PgUb$ suppressions are observed to increase with collision centrality.
The suppression of \PgUb\ is stronger than that of \PgUa\ in all centrality ranges,
including the most peripheral bin.  It should be noted that
this bin (50--100\%) is rather wide and mostly populated by
more central events (closer to 50\%).
For this most peripheral bin the \PgUa nuclear modification factor is 
$1.01 \pm 0.12\,(\text{stat.}) \pm 0.22\,(\text{syst.})$,
while for the most central bin (0--5\%) \raa is $0.41 \pm 0.04\,(\text{stat.}) \pm 0.07\,(\text{syst.})$
indicating a significant suppression.
The observed $\PgUn$ yields contain contributions from decays of heavier bottomonium states and,
thus, the measured suppression is affected by the dissociation of these states.
This feed-down contribution to the $\PgUa$ state was measured to be of the order of 50\%~\cite{Affolder:1999wm, lhcbfeeddown}, albeit in different kinematic ranges than used here. 
These results indicate that the directly produced $\PgUa$ state is not significantly suppressed, however quantitative conclusions will require precise estimations of the feed-down contribution matching the phase space of the suppression measurement.

In addition to QGP formation, differences between quarkonium production yields
in \PbPb and \pp collisions can also arise from cold-nuclear-matter
effects~\cite{Vogt:2010aa}.  However, such effects should have a small impact on the double ratios reported here.
Initial-state nuclear effects are expected to affect similarly each of the three \PgU states,
thereby canceling out in the ratio.
Final-state ``nuclear absorption'' becomes weaker with increasing energy~\cite{Lourenco:2008sk} and is expected to be negligible at the LHC~\cite{Lin:2000ke}.
Future high-statistics heavy-ion, proton-proton, and proton-nucleus runs at the LHC will provide further quarkonium measurements, which will help to disentangle cold-nuclear from hot-medium effects and to attain a more thorough characterization of the properties of the produced medium.

In conclusion, the first observation of the sequential suppression of the \PgUn\ states in heavy-ion collisions has been reported, in $\sqrtsnn = 2.76$\TeV \PbPb collisions by the CMS experiment at the LHC, extending the previous CMS bottomonium measurements~\cite{prl,HIN-10-006}. The \PgUb\ and \PgUc\ resonances are suppressed with respect to the \PgUa state, with a significance exceeding $5\,\sigma$. The nuclear modification factors for the \PgUn\ states were also measured, with the individual \PgUa, \PgUb, and \PgUc\ states  suppressed by factors of about 2, 8, and larger than 10, respectively.

We congratulate our colleagues in the CERN accelerator departments for the excellent performance of the LHC machine. We thank the technical and administrative staff at CERN and other CMS institutes, and acknowledge support from: BMWF and FWF (Austria); FNRS and FWO (Belgium); CNPq, CAPES, FAPERJ, and FAPESP (Brazil); MES (Bulgaria); CERN; CAS, MoST, and NSFC (China); COLCIENCIAS (Colombia); MSES (Croatia); RPF (Cyprus); MEYS (Czech Republic); MoER, SF0690030s09 and ERDF (Estonia); Academy of Finland, MEC, and HIP (Finland); CEA and CNRS/IN2P3 (France); BMBF, DFG, and HGF (Germany); GSRT (Greece); OTKA and NKTH (Hungary); DAE and DST (India); IPM (Iran); SFI (Ireland); INFN (Italy); NRF and WCU (Korea); LAS (Lithuania); CINVESTAV, CONACYT, SEP, and UASLP-FAI (Mexico); MSI (New Zealand); PAEC (Pakistan); MSHE and NSC (Poland); FCT (Portugal); JINR (Armenia, Belarus, Georgia, Ukraine, Uzbekistan); MON, RosAtom, RAS and RFBR (Russia); MSTD (Serbia); SEIDI and CPAN (Spain); Swiss Funding Agencies (Switzerland); NSC (Taipei); TUBITAK and TAEK (Turkey); NASU (Ukraine); STFC (United Kingdom); DOE and NSF (USA).

\bibliography{auto_generated}
\ifthenelse{\boolean{cms@external}}{}{
\appendix
\section{Centrality-dependent results (tabulated format)\label{app:suppMat}}

\begin{table}[!h]
\centering
\topcaption{Double ratio versus centrality.} 
\begin{tabular}{r|c}
\hline
$\npart$ (centrality)  &  $\frac{\PgUb/\PgUa|_{\PbPb}}{\PgUb/\PgUa|_{\pp}} \pm \text{stat.}\pm \text{syst.} \pm \text{global}$  \\
\hline
381 ( 0--5\%)      &  $0.270 \pm 0.146 \pm 0.031 \pm 0.032$ \\
329 ( 5--10\%)     &  $0.102 \pm 0.189 \pm 0.042 \pm 0.012$ \\
261 (10--20\%)     &  $0.138 \pm 0.108 \pm 0.041 \pm 0.017$ \\
187 (20--30\%)     &  $0.428 \pm 0.124 \pm 0.039 \pm 0.051$ \\
130 (30--40\%)     &  $0.344 \pm 0.138 \pm 0.027 \pm 0.041$ \\
 86 (40--50\%)     &  $0.420 \pm 0.220 \pm 0.048 \pm 0.050$ \\
 22 (50--100\%)    &  $0.304 \pm 0.154 \pm 0.040 \pm 0.037$ \\
\hline
\end{tabular}
\label{tab:doubleratio}
\end{table}

\begin{table}[!h]
\centering
\topcaption{\PgUa\ and \PgUb\ nuclear modification factors, \raa, versus centrality.} 
\begin{tabular}{r|c|c}
\hline
$\npart$ (centrality)  &  $\PgUa\,\raa \pm \text{stat.}\pm \text{syst.} \pm \text{global}$  &  $\PgUb\,\raa \pm \text{stat.}\pm \text{syst.}  \pm \text{global}$  \\
\hline
381 ( 0--5\%)      &  $0.411 \pm 0.043 \pm 0.048 \pm 0.056$ &   $ 0.111 \pm 0.061 \pm 0.020 \pm 0.024$ \\
329 ( 5--10\%)     &  $0.432 \pm 0.048 \pm 0.046 \pm 0.059$ &   $ 0.044 \pm 0.060 \pm 0.019 \pm 0.009$ \\
261 (10--20\%)     &  $0.484 \pm 0.040 \pm 0.049 \pm 0.066$ &   $ 0.068 \pm 0.053 \pm 0.022 \pm 0.014$ \\
187 (20--30\%)     &  $0.614 \pm 0.053 \pm 0.075 \pm 0.084$ &   $ 0.260 \pm 0.079 \pm 0.043 \pm 0.055$ \\
130 (30--40\%)     &  $0.681 \pm 0.069 \pm 0.085 \pm 0.093$ &   $ 0.237 \pm 0.098 \pm 0.039 \pm 0.050$ \\
 86 (40--50\%)     &  $0.590 \pm 0.096 \pm 0.086 \pm 0.080$ &   $ 0.251 \pm 0.138 \pm 0.050 \pm 0.054$ \\
 22 (50--100\%)    &  $1.005 \pm 0.121 \pm 0.176 \pm 0.137$ &   $ 0.300 \pm 0.157 \pm 0.069 \pm 0.064$ \\
\hline
\end{tabular}
\label{tab:raa}
\end{table}

}
\cleardoublepage \appendix\section{The CMS Collaboration \label{app:collab}}\begin{sloppypar}\hyphenpenalty=5000\widowpenalty=500\clubpenalty=5000\textbf{Yerevan Physics Institute,  Yerevan,  Armenia}\\*[0pt]
S.~Chatrchyan, V.~Khachatryan, A.M.~Sirunyan, A.~Tumasyan
\vskip\cmsinstskip
\textbf{Institut f\"{u}r Hochenergiephysik der OeAW,  Wien,  Austria}\\*[0pt]
W.~Adam, E.~Aguilo, T.~Bergauer, M.~Dragicevic, J.~Er\"{o}, C.~Fabjan\cmsAuthorMark{1}, M.~Friedl, R.~Fr\"{u}hwirth\cmsAuthorMark{1}, V.M.~Ghete, J.~Hammer, N.~H\"{o}rmann, J.~Hrubec, M.~Jeitler\cmsAuthorMark{1}, W.~Kiesenhofer, V.~Kn\"{u}nz, M.~Krammer\cmsAuthorMark{1}, I.~Kr\"{a}tschmer, D.~Liko, I.~Mikulec, M.~Pernicka$^{\textrm{\dag}}$, B.~Rahbaran, C.~Rohringer, H.~Rohringer, R.~Sch\"{o}fbeck, J.~Strauss, A.~Taurok, W.~Waltenberger, G.~Walzel, E.~Widl, C.-E.~Wulz\cmsAuthorMark{1}
% ...
\vskip\cmsinstskip
\dag:~Deceased\\
1:~~Also at Vienna University of Technology, Vienna, Austria\\
% ...

\end{sloppypar}
\end{document}